\documentclass[traditabstract]{aa} 
\usepackage{graphicx}
\usepackage{longtable}
\usepackage{rotating}
\usepackage{natbib}
\usepackage[usenames]{color}
\usepackage{multirow}
\usepackage{url}
\usepackage{ulem}
\bibpunct{(}{)}{;}{a}{}{,} 
\newcommand{\ratio} {N({\rm H}_2) / I_{\rm CO(1-0)}}
\newcommand{\ratioo} {N({\rm H}_2) / I_{\rm CO}}

\newcommand{\kms}   {{\rm \  km \  s^{-1}}}
\newcommand{\K}   {{\rm \  K}}

\newcommand{\Xunit} {\,{\rm cm^{-2}/(K\kms)}}

\newcommand{\micron}{\mbox{$\mu$m}}   
\newcommand{\HI}{\ion{H}{i}}   
\newcommand{\CII}{[\ion{C}{ii}]}   
\newcommand{\HII}{\ion{H}{ii}}   
   
\newcommand{\msun}{M$_{\odot}$}   
\newcommand{\msol}{M$_{\odot}$}   
\newcommand{\moyr}{M$_{\odot}$yr$^{-1}$}   

\usepackage{txfonts}

\begin{document}
   \title{A detailed view of a Molecular Cloud in the far outer disk of M~33}
   \titlerunning{A Molecular Cloud in the far outer disk of M~33}
   \subtitle{Molecular Cloud Formation in M~33 \thanks{Based on observations carried out with the IRAM Plateau de Bure Interferometer. IRAM is supported by INSU/CNRS (France), MPG (Germany) and IGN (Spain).}}

   \author{J. Braine \inst{1,2} \and P. Gratier \inst{3} \and  Y. Contreras \inst{4} \and K.F. Schuster  \inst{3}   \and N. Brouillet\inst{1,2} 
        }

   \institute{Univ. Bordeaux, Laboratoire d'Astrophysique de Bordeaux, F-33270, Floirac, France.\\
             \email{braine@obs.u-bordeaux1.fr}
        \and
  CNRS, LAB, UMR 5804, F-33270, Floirac, France
         \and
            IRAM, 300 Rue de la Piscine, F-38406 St Martin d'H\`eres, France   
         \and  Univ. de Chile, Santiago, Chile
        }

   \date{}

  \abstract{ 
  The amount of H$_2$ present in spiral galaxies remains uncertain, particularly in the dim outer regions and in low-metallicity environments.  We present high-resolution CO(1--0) observations with the Plateau de Bure interferometer of the most distant molecular cloud in the local group galaxy M~33.  The cloud is a single entity rather than a set of smaller clouds within the broad beam of the original single-dish observations.  The interferometer and single-dish fluxes are very similar and the line widths are indistinguishable, despite the difference in beamsize.  At a spatial resolution of 10~pc, beyond the optical radius of the M~33, the CO brightness temperature reaches 2.4~Kelvins.  A virial mass estimate for the cloud yields a mass of $4.3 \times 10^4$ \msun\ and a ratio $\ratio \simeq 3.5 \times 10^{20} \Xunit$.  While no velocity gradient is seen where the emission is strong, the velocity is redshifted to the extreme SW and blue-shifted to the far NE.  If the orientation of the cloud is along the plane of the disk (i.e. not perpendicular), then these velocities correspond to slow infall or accretion.  The rather modest infall rate would be about $2 \times 10^{-4}$\moyr.}

\maketitle

\section{Introduction}

How much molecular gas is present in spiral galaxies?  What determines whether or not a molecular cloud will form in a given place and time?  Do all molecular clouds form stars and, if so, at what rate? These are central questions to understanding where and why stars form in galaxies yet they are difficult to answer.  One of the major difficulties is the absence of direct observation of cool H$_2$.  The standard tracer of molecular gas, the CO molecule, which is the most abundant molecule with a rotational dipole moment, is clearly a poor tracer of H$_2$ in very low-metallicity
environments \citep{Rubio91,Bolatto08} and contested in other situations, particularly in the cool outer parts of spirals \citep{Pfenniger94a,Sodroski95}.  CO emission has been detected in only a very few galaxies beyond the nominal optical radius $R_{25}$ \citep[e.g.][]{Digel94,Brand87} for the Galaxy and \citet{Braine04b} and \citet{Braine07} for other galaxies.  

The Local Group spiral M~33, $R_{25} = 30.8'  \approx 7.5$ kpc, is an ideal candidate to better study the above issues because it is very nearby, such that molecular clouds can be detected individually, with a classical spiral disk structure with no geometrical uncertainties, and an only slightly subsolar metallicity \citep[$\sim$factor 2,][]{Magrini10}.  The observations presented here are part of a series by our group \citep{Gardan07, Gratier10, Gratier12,Kramer10,Braine10b} on the gas and dust of M~33 with similar goals to work by {\it e.g.} \citet{Wilson97}, \citet{Engargiola03}, \citet{Rosolowsky07}, or \citet{Bigiel10}.  In this work, we present high resolution observations with the Plateau de Bure interferometer of the very distant cloud detected  as M33-18 by \citet{Braine10a}, just beyond $R_{25}$ and the most distant detected to date.
This is roughly equivalent to twice the solar circle distance.

Being able to spatially resolve clouds is important to determine whether the weak emission in the outer disks of spirals is due to the very low brightness temperatures of the clouds or to their scarcity in a telescope beam encompassing  an area much greater than that of a molecular cloud.  Increasing the resolution enables us to study the cloud itself to determine whether it is a single cloud and to estimate the mass necessary to gravitationally bind it together.  This is the goal of the current work.  

\section{Observations and data reduction}

The most distant molecular cloud to date in the disk of M~33 has now been observed at high resolution 
in the CO(1--0) line.
The data were acquired in early 2009 with the interferometer on the Plateau de Bure operated by the Institut de RadioAstronomie Millim\'etrique in the `C' and `D' configurations. A single field was observed centered on RA (J2000) 01:34:12.5 and Dec (J2000) 31:10:32.

\begin{figure}[!h!]
\begin{flushleft}
\includegraphics[angle=0,width=8.5cm]{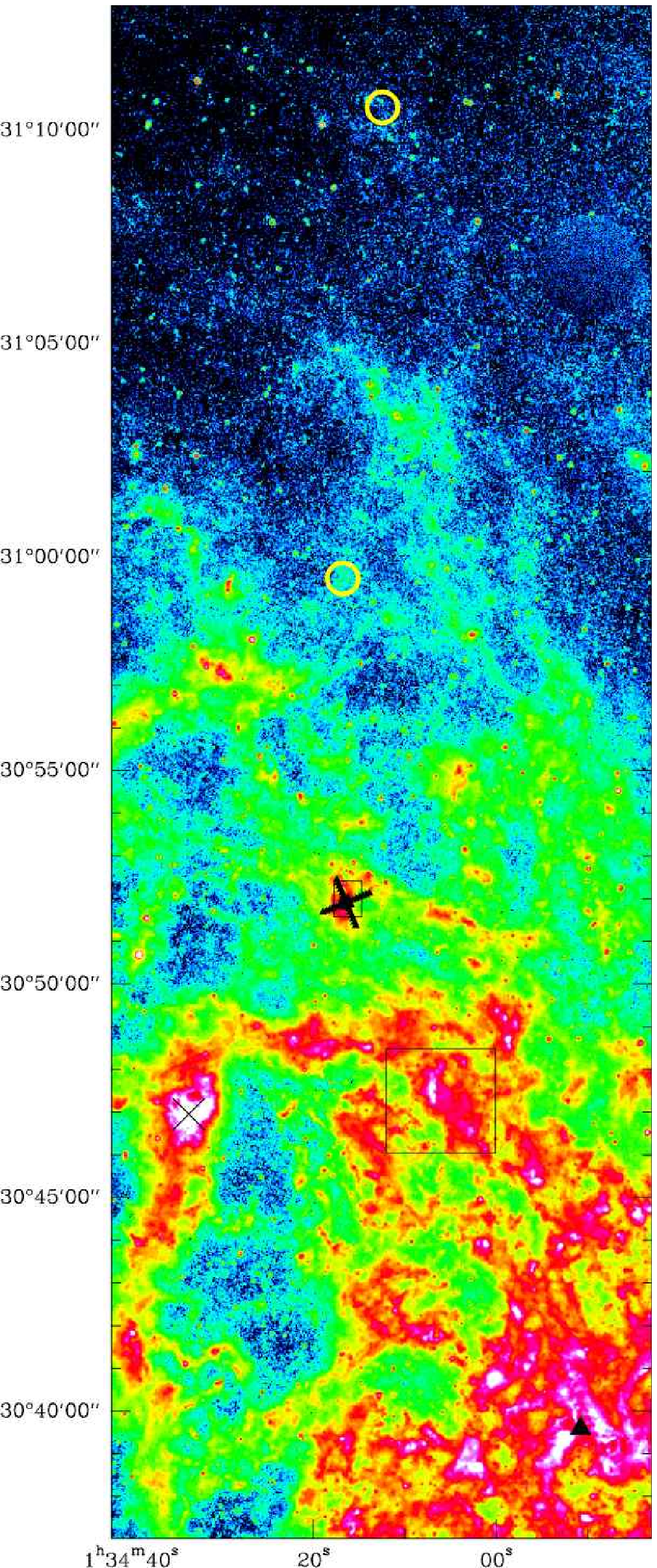}
\caption{\label{b691_fig1} North-eastern part of M~33 in the IRAC
8~\micron\ band \citep{Verley07}.  
The cloud we present here is in the small circle at the top center of the image, 
representing the primary beam of the Plateau de Bure interferometer at the frequency of CO(1--0).
The "Lonely Cloud", a fairly CO-strong inter-arm GMC \citep{Gardan07b,Braine10a}, is also indicated 
with a small circle.  
The small rectangle at Dec. 30:52 and the small
triangles forming an 'X' in and around the rectangle indicate the
individual positions of the \CII\ HIFI observations of the \HII\ region BCLMP~691 presented in Braine et al (2012).  
The large lower rectangle shows the region around BCLMP~302 observed with PACS by \citet{Mookerjea11}.
  The center of M~33 is at the position of
  the triangle to the lower right of the image of M~33.  The giant
  \HII\ region NGC~604 can be seen as the very bright region to the
  left around Dec. 30:47 and is marked with a $\times$.}
\end{flushleft}
\end{figure}

Calibration was straightforward and standard routines within CLIC were used.
3C454.3 and MWC~349 were observed once at the beginning of each track respectively as bandpass and flux flux calibrators, 0234+285 was used as phase and amplitude calibrator  and was observed every 12 minutes for the remainder of the track. The calibration uncertainties at 115~GHz at the PdBI are estimated to be 10\%. 

Datacubes were made at 4 different spatial resolutions by varying the weighting scheme applied to the visibilities.
Tapering at 50 and 90 meters with natural weighting yielded a beam of HPBW $6.65\arcsec \times 5.46\arcsec$ (PA $99^\circ$) and $4.50\arcsec \times 3.49\arcsec$ (PA $114^\circ$) respectively.  Pure natural weighting with no taper, in order to maximize point source sensitivity, produced a beam of HPBW $3.58\arcsec \times 2.67\arcsec$ (PA $119^\circ$) and robust weighting gave a resolution of $2.74\arcsec \times 2.40\arcsec$ (PA $105^\circ$).  These angular resolutions correspond to linear resolutions of 10 -- 25 pc for an assumed distance to M~33 of 840 kpc \citep{Galleti04}.  The initial spectral resolution was $0.1\kms$ (39~kHz) but in all cases two channels were combined such that the final channel separation was $0.2 \kms$.  Each cube has 76 channels.  All image processing, including deconvolution, was done in the MAPPING environment (see http://www.iram.fr/IRAMFR/GILDAS).

Interferometric observations always have a complex beam pattern so deconvolution (``cleaning") is necessary and is often the most delicate part of the data reduction.  The dataset with the lowest angular resolution has the highest brightness sensitivity and is thus most sensitive to extended structures.  The ``dirty" cubes produced as per the previous paragraph were first cleaned with the default clean box, which covers much of the central region and yields an unbiased indication of where signal is present.  For the 22 channels which appeared to show emission, polygons were drawn to indicate the regions to clean and this enabled us to further improve the imaging.  The Clark clean algorithm was used with a maximum of 10000 iterations and 500 major cycles except for one channel where we used 1000 iterations.  The loop gain in all cases was 0.1.
The higher resolution cubes were cleaned in the same way but the polygons from the low-resolution data were used for the cleaning in order to avoid the bias of following positive noise at high angular resolution.


\section{Morphology, dynamics, and mass of the distant outer disk cloud}

\begin{figure}[!h!]
\begin{flushleft}
\includegraphics[angle=0,width=8.5cm]{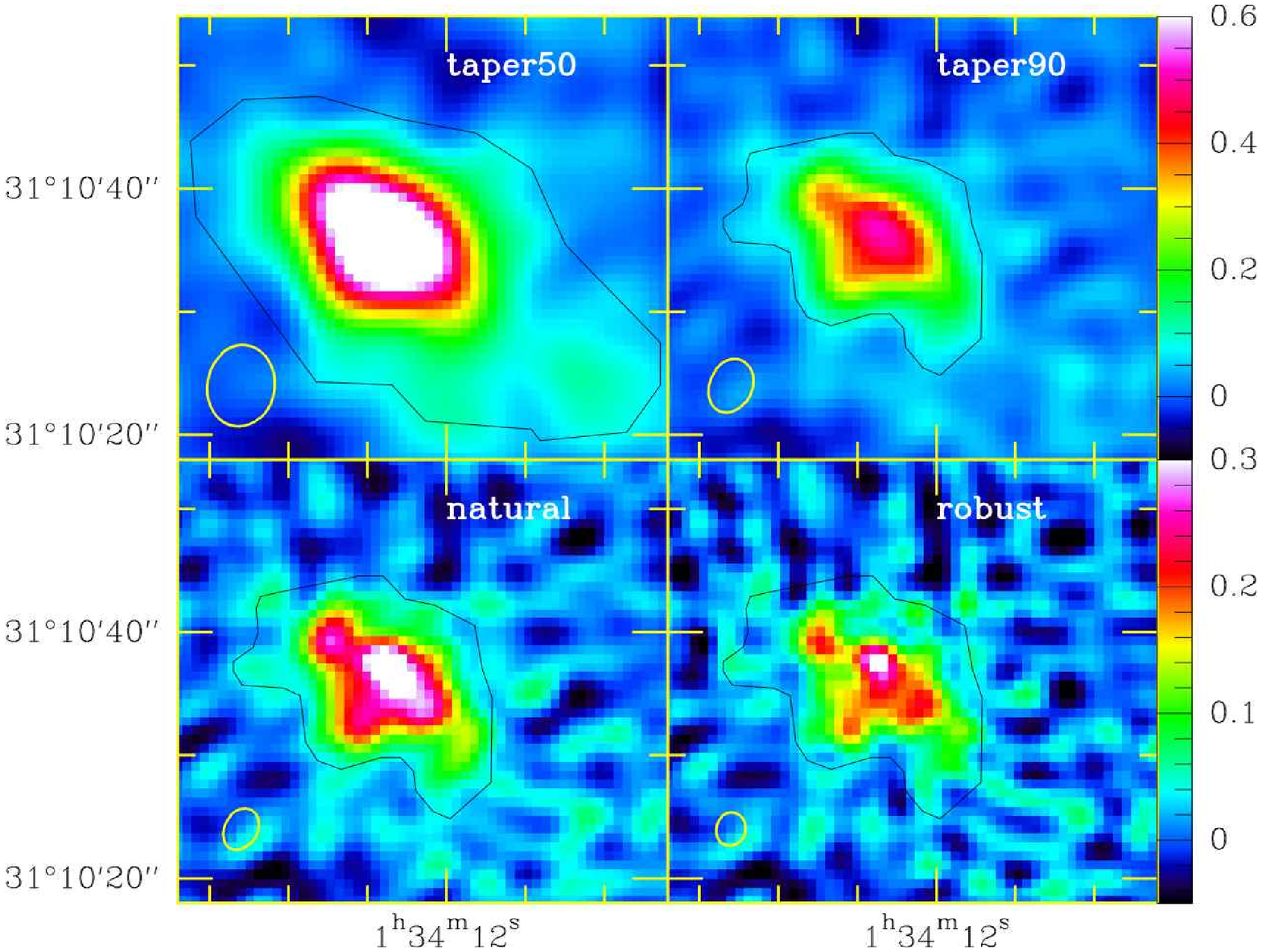}
\caption{\label{m33c_fig_mom0} Morphology of the outer disk cloud; shown are the moment zero or integrated intensity maps made with the tapered, natural, and robust weighting.  The beam sizes are indicated as ellipses in the lower left corner of each map.
The color scales are shown to the right for the tapered and untapered images separately and are in 
units of Jy/beam. The  extended emission contour is shown in the upper left box and the cloud contour is drawn in the other panels.  These two contours are used to compute the fluxes given in Table~\ref{tab.flux}.}
\end{flushleft}
\end{figure}

As can be seen in Fig.~\ref{b691_fig1}, the cloud presented here is much further out in the disk of M~33 than
others detected up to now.  Such clouds are rare and difficult to find \citep[see also cloud M33-20 in][]{Braine10a}.
Figure~\ref{m33c_fig_mom0} shows the integrated intensity (moment zero) maps at the four resolutions.  
We have drawn what we consider to be the cloud outline on the panels in Fig.~\ref{m33c_fig_mom0} 
except for the upper left panel which shows the lowest resolution -- but highest brightness sensitivity -- data
where we draw a contour including possible extended emission, a sort of maximum cloud size derived from
the sum of the polygons used for cleaning the data.
Table~\ref{tab.flux} provides the integrated fluxes within these contours.  

\citet{Braine10a} measured an integrated intensity of 0.61 K$\kms$, or 2.9 Jy$\kms$, in the CO(1--0) line at $21\arcsec$ resolution with the IRAM 30 meter telescope but the 
pointing center was 3 arcseconds south of the cloud peak (which was not known before these observations).  As can be seen from Table~1, very little if any flux is resolved out by the present interferometric measurements.  
The 'cloud' contour in Table~1 has an area of half that of the 30meter telescope beam and
the 'extended' emission contour is about 1/3 larger than the 30m beamsize.  

\begin{figure}[!h!]
\begin{flushleft}
\includegraphics[angle=0,width=8.5cm]{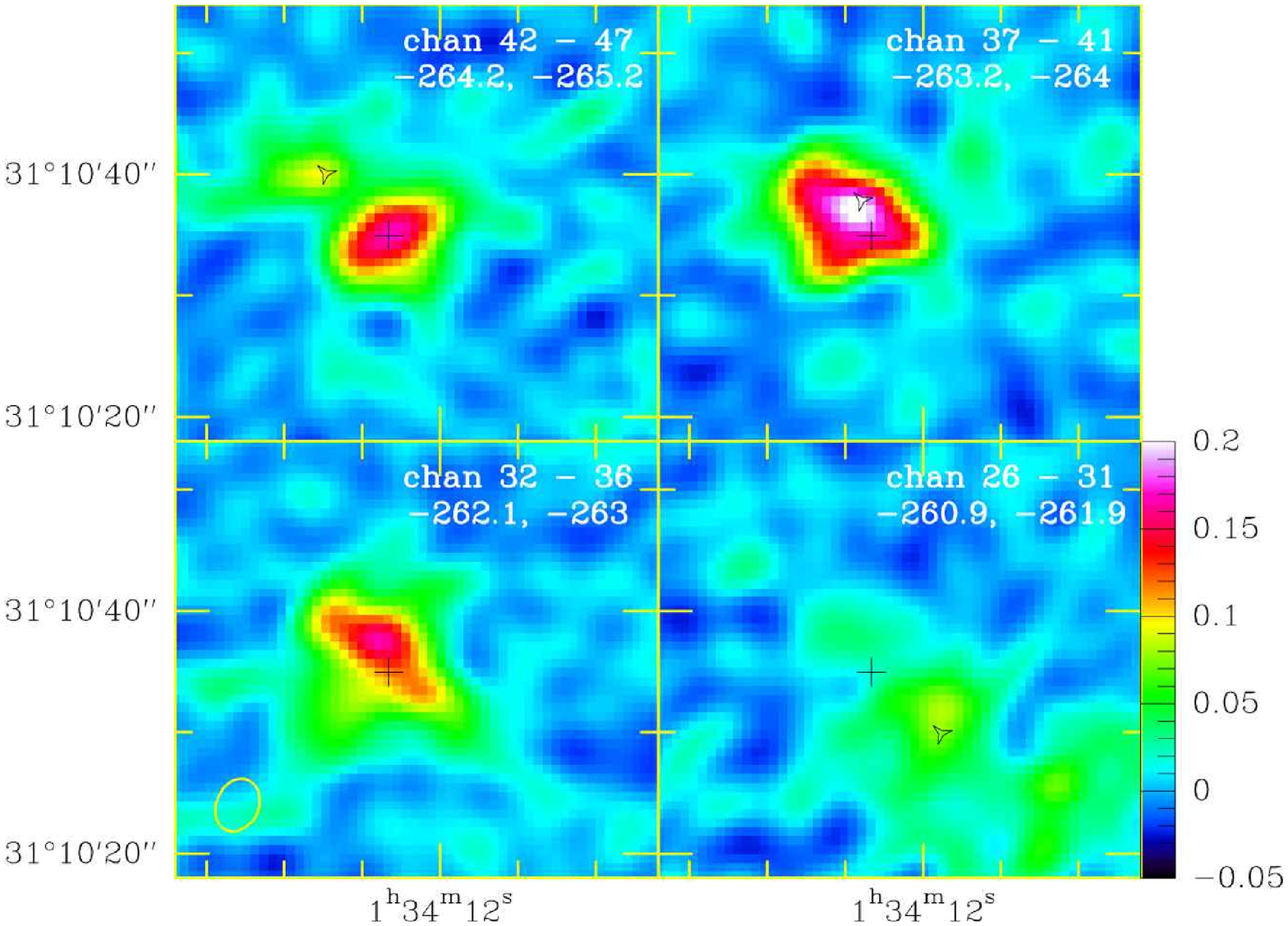}
\caption{\label{m33c_fig_mom0c} Velocity structure of the outer disk cloud at the resolution of the map tapered at 90 meters. The cross  marks RA 01:34:12.3, Dec 31:10:35 and is what we later refer to as the (0,0) position; the triangles in the panels indicate the high-velocity, peak, and low-velocity components shown in Fig.~\ref{m33c_spec}. }
\end{flushleft}
\end{figure}

Figure~\ref{m33c_fig_mom0c} shows the structure of the source as a function of velocity, integrating over about $1 \kms$ and covering the velocity extent of the CO emission.  While almost all of the flux is contained within a $3\kms$ interval, there is a separate patch of weak emission at lower rotation velocity detected to the southwest,
seen in the lower right panel of Fig.~\ref{m33c_fig_mom0c}. 

\begin{figure}[!h!]
\begin{flushleft}
\includegraphics[angle=0,width=8.5cm]{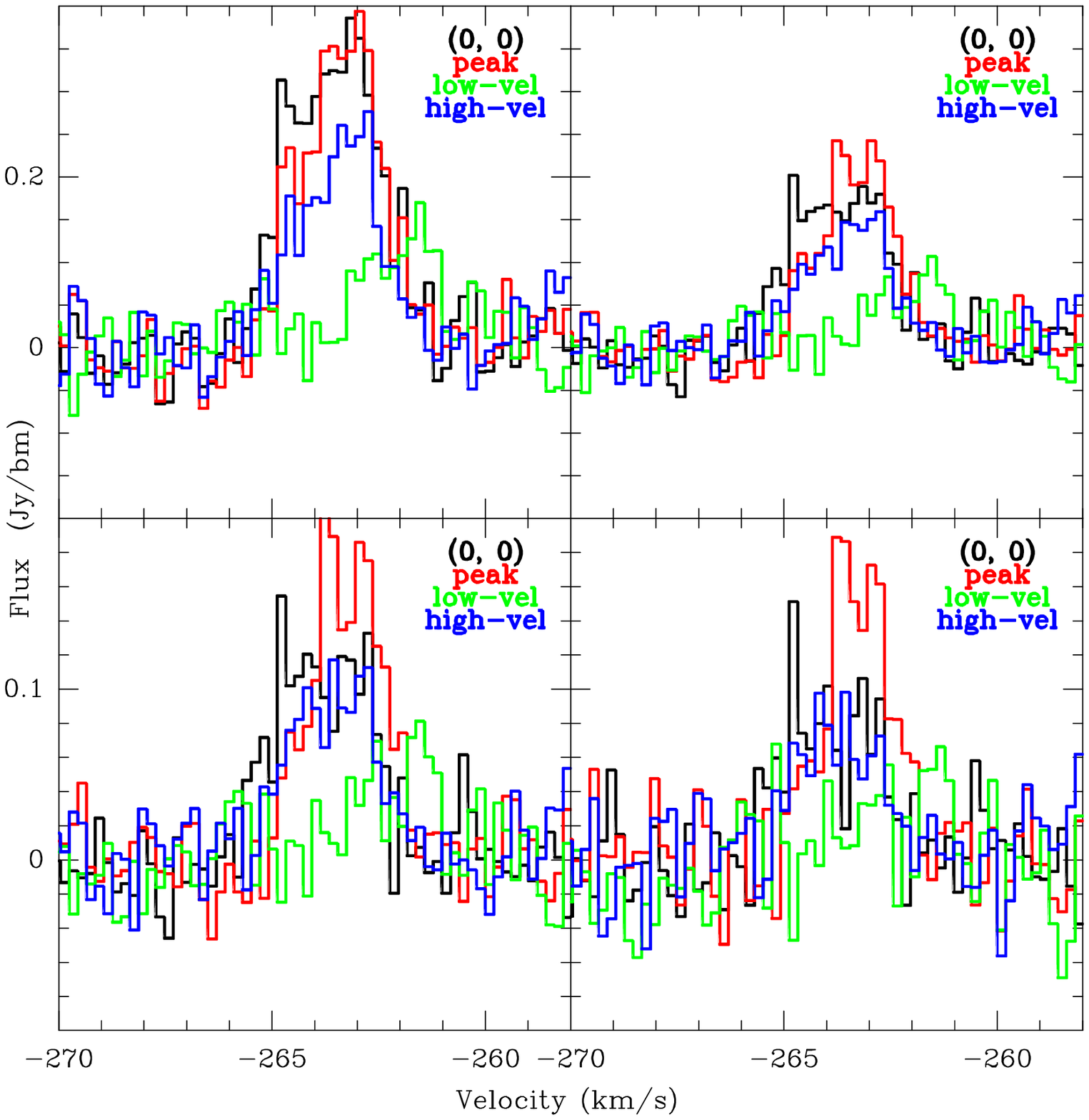}
\caption{\label{m33c_spec} Spectra as a function of position in the cloud.  The (0,0) position refers to the cross in Fig.~\ref{m33c_fig_mom0c}.  
The ``peak" position corresponds to the triangle in the upper right panel of Fig.~\ref{m33c_fig_mom0c}.
The ``high-vel" position corresponds to the triangle in the upper left panel of Fig.~\ref{m33c_fig_mom0c}.
The ``low-vel" position corresponds to the triangle in the lower right panel of Fig.~\ref{m33c_fig_mom0c}.
The $y-$scale varies because the spectra are expressed in Jy/beam so the larger beams have greater fluxes.  
The tapered and untapered spectra are shown in the same order as in Fig.~\ref{m33c_fig_mom0}. }
\end{flushleft}
\end{figure}

Figure~\ref{m33c_spec} shows the spectra at four positions in the cloud at the four resolutions.  The full line width at zero intensity is $4.4\kms$ and at half power the line width is $2.4 - 2.6 \kms$ measured directly on the integrated spectrum.  The line is 12 or 13 channels wide, depending on the exact region over which the spectrum is summed.  Interestingly, the line width does not change with the angular resolution -- the highly tapered data have the same linewidths as the high resolution cubes.  Also, the line width changes little with position over the cloud, at least with the $\sim10$~pc resolution presented here.  The line width measured with the 30 meter telescope is the same: $\Delta {\rm v} = 2.5 \pm 0.2 \kms$ \citep{Braine10a}.

Even at the highest angular resolution, the cloud does not break down into smaller distinct entities, with the possible exception of the ``low-vel" region seen to the south-west.  The cloud covers about about $250$ square arcsec or 4000 pc$^2$.  Unlike \citet{Braine10a}, here we are able to measure the size of the cloud.  Following \citet{Solomon87}, we equate $R_e = \sqrt{A/\pi}$, and 
$S_{S87} = \sqrt{\pi} R_e / 3.4 = \sqrt{A}/3.4$.  
Let us note however, that while \citet{Solomon87} delimited clouds at a
brightness of several Kelvin, the cloud size we measure extends close to the zero-intensity level.  Adopting their parameters for consistency, $M_{vir} = 3 f_p \sqrt{A} \sigma_{\rm v}^2 /3.4 G$.  With $f_p=2.9$ and 
$\sigma_{\rm v} = \Delta {\rm v} /2.35$, we estimate a mass of $4.26 \times 10^4$ \msun.  This is higher by about 30\% 
than the previous estimate by \citet{Braine10a} based on the single-dish spectra.
The flux within the 4000 pc$^2$ region in the robust map is 3.18 Jy$\kms$.  The conversion ratio 
$$ \ratio = \frac{M_{vir} \, f_h} {2 \, m_p \, D^2 \, \Omega} \, \, \frac{2 \, k_b\, \nu^2 \, \Omega}{c^2 \, S \, \Delta {\rm v}} $$
where the first term on the right is the H$_2$ column density and the second term is I$_{\rm CO}^{-1}$;
$D$ is the distance to M~33, $m_p$ is the proton mass, $f_h$ is the hydrogen fraction by mass (0.73), 
$\Omega$ is the solid angle occupied by the cloud, and 
$S \Delta {\rm v}$ is the velocity integrated flux.  For the robust map flux of $S \Delta {\rm v} =$3.18 Jy$\kms$, 
$\ratio = 3.5 \times 10^{20} \Xunit$.

If instead we use the rather uncertain extended emission contour, then the mass becomes $7.07 \times 10^4$ \msun\ and  
$\ratio = 4.4 \times 10^{20} \Xunit$.  Our estimates of $\ratioo$ are at the upper end of what \citet{Leroy11} estimated for M~33.

\begin{table}[h]
\caption[]{ Fluxes for the extended emission and cloud contours for each angular resolution.
The first figure in each column is the flux measured before primary beam correction and the second after correction.
We estimate the statistical uncertainties (including the uncertainty in the cloud contour) to be $\sim 15$\%.}
\label{tab.flux}
\begin{tabular}{lrrrr}   
\hline  
zone & taper50 & taper90 &natural & robust \\
 & Jy$\kms$ & Jy$\kms$ & Jy$\kms$ & Jy$\kms$ \\
\hline  
cloud & 2.5, 2.7 & 2.8, 3.0 & 2.9, 3.1 & 3.0, 3.2 \\
extended & 3.4, 3.9 & 3.6, 4.1 & 3.6, 4.1 & 3.7, 4.2 \\
\end{tabular}   
\end{table}   

\begin{figure*}[!h!]
\begin{flushleft}
\includegraphics[angle=0,width=18cm]{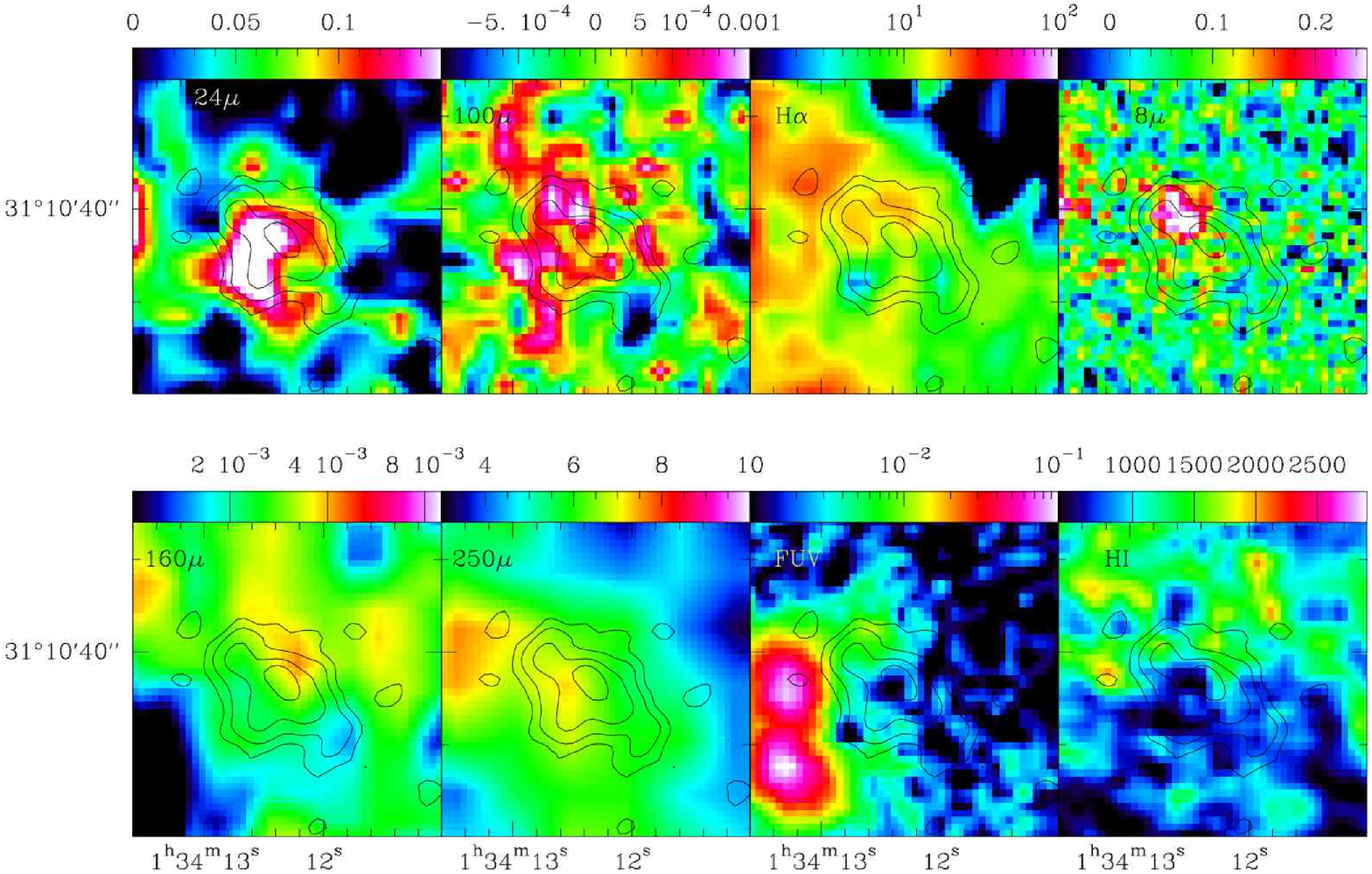}
\caption{\label{maps} Comparison of the CO(1--0) emission at the resolution of the natural weighting -- contours at 0.05, 0.1, 0.2, 0.3 Jy$\kms$ -- with the emission in the Far-IR, H$\alpha$ (arbitrary units), Far-UV (arbitrary units), \HI, and 8\micron PAH bands.  From left to right (top row) units are: MJy/sr, Jy/pixel, arbitrary, MJy/sr and (bottom) Jy/pixel, MJy/sr, arbitrary, and integrated \HI\ line intensity in $\K \kms$. 
Labels indicating the wavebands are provided in the upper left corners of each panel and the color wedge is shown above.  The \HI\ data is at the full $5\arcsec$ resolution of the VLA mosaic described in \citet{Gratier10}.}
\end{flushleft}
\end{figure*}

In the highest resolution map, not tapered and with robust weighting (lower right panel of Fig.~2), the peak brightness temperature reaches 2.4~K.  Although not extreme, a brightness temperature of 2.4~K at 10 pc resolution  is actually quite high, particularly considering the weak radiation field surrounding the cloud.  The emission in the GALEX Far-UV band just to the East of the cloud (see Fig.~\ref{maps}) suggests a Far-UV field several times the value in the solar neighborhood, only some of which will be felt by the cloud, depending on the  solid angle occupied by the cloud seen from the stars.   

Figure~\ref{maps} shows the CO emission in comparison with other wavebands.  While IR or H$\alpha$ emission at or near the position of the cloud is present, it is typically 2 orders of magnitude weaker than towards  star-forming
regions in the disk of M~33 \citep[e.g.][]{Mookerjea11,Braine12} although the latter have only slightly 
stronger CO emission.  Using Eq. 9 of \citet{Calzetti07}, the 24 $\mu$m, which is the waveband which most clearly
shows star formation associated with the molecular cloud, yields a star formation rate (SFR) of 4 \msol\ Myr$^{-1}$.
This is equivalent to a conversion rate of gas into stars of M(H$_2$)/ SFR $\sim 10^{10}$ years, close to an 
order of magnitude slower than in spiral galaxies in general \citep[e.g.][]{Kennicutt98b} and M~33 in particular \citep{Gardan07,Gratier10}.  

The CO(2--1) emission of this cloud, while not mapped, is twice as strong
in brightness units \citep{Braine10a} towards the center; given the cloud size, this implies that the gas kinetic  temperature and density are sufficient to efficiently excite the $J=2-1$ transition \citep[see e.g. discussion 
in App. 1 of][]{Braine_bs1}.  Since little or no flux is resolved out by the interferometer, we can smooth to the 
$\sim11"$ beam of the 30 meter telescope in CO(2--1); the observations presented here confirm a 
CO(2--1)/CO(1--0) ratio close to unity at a scale of $11"$. 
The neighboring UV emission sources cannot heat the interior of the cloud 
due to the high optical depth. The level of star formation, as estimated using the 24 or 8~\micron\ emission, 
is much lower than in many other clouds in M~33 with similar CO emission.  

{\it How does this extreme outer disk cloud compare with other GMCs?}
\citet{Gratier12} observed a large sample of GMCs with an angular resolution of 12$"$ and found 
consistently larger line widths.  This is not due to the angular resolution, however, because the line width at
21$"$ resolution \citep{Braine10a} is the same as that found here.  \citet{Gratier12} compared the M~33
cloud sample with Milky Way clouds and found that the M~33 clouds had smaller line widths than Milky Way 
inner disk clouds -- such that the line width of our outer disk cloud (M33-18) is far below that of the \citet{Solomon87}
sample of first quadrant molecular clouds, in a completely empty region in their Fig. 1.
Perhaps more surprisingly, the \citet{Bigiel10} clouds in M~33 (considerably closer to the center, near RA 01:34, Dec 30:55) and Large Magellanic Cloud sample also have broader lines, whether 
for star-forming clouds or not \citep{Kawamura09}.  
On the other hand, the \citet{Digel94} sample of Milky Way outer disk clouds is very similar, with slightly narrower
lines and somewhat smaller sizes.  The very distant cloud 2 found by \citet{Digel94} and studied further by \citet{Ruffle07} is similar in CO luminosity and line width to M33-18 although smaller both in Virial mass and size, using the average of the distances used by \citet{Ruffle07}.  \citet{Brand95} also find that the Milky Way outer disk clouds have 
linewidths about a factor 2 smaller than inner disk clouds of the same size. The CO brightness at 10pc resolution is also similar to far outer Galaxy molecular clouds \citep{Brand87,Digel94}.

\section{Gas inflow onto the molecular cloud}

\begin{figure}[!h!]
\begin{flushleft}
\includegraphics[angle=0,width=8.8cm]{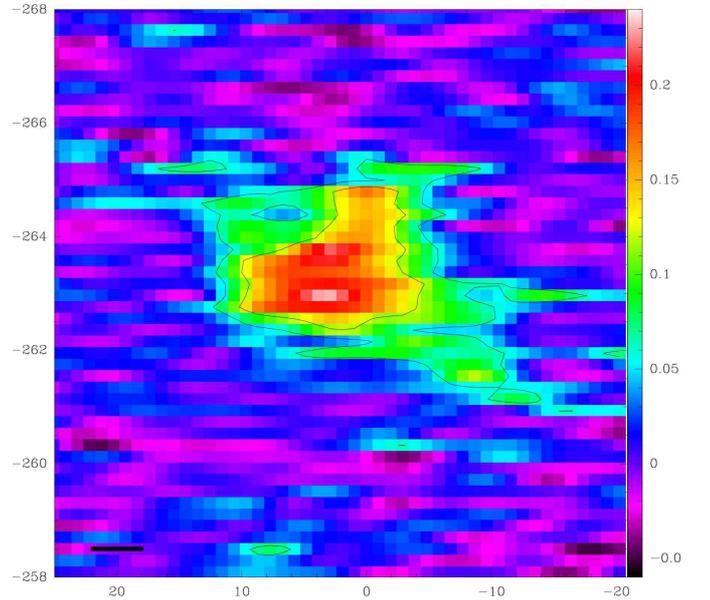}
\caption{\label{pv} PV-diagram of the outer disk cloud along a cut with a position angle of 49$^\circ$
East of North, made with the datacube tapered at 90meters. 
The Y-axis indicates recession velocity in $\kms$ from slower rotation ($-258 \kms$) to faster rotation velocities.
The X-axis indicates the distance from the center 
in arcseconds, with negative values being to the south-west.  The color scale is in Jy per beam.
The bar to the lower left indicates $4"$, as the beam measures $4.50 \times 3.49"$.}
\end{flushleft}
\end{figure}

Although the \HI\ emission at the position of the cloud is not particularly strong, the cloud is part of an arm-like structure seen in \HI\ and it follows the same orientation.  
Looking at Fig.~\ref{m33c_fig_mom0c} more closely, we see that the 
blue-shifted emission is to the NE and the red-shifted region to the SW.  Assuming that the spiral arms are 
trailing, the West side of M~33 is the near side and thus, as long as the cloud extensions are in the plane of the galaxy (reasonable given that they follow the arm orientation), the blue and red-shifted emission both correspond to infall towards the center of the molecular cloud.  
Figure~\ref{pv} shows the position-velocity (PV) diagram for the cloud, following the general orientation of the HI arm \citep[see Fig.~5 in ][]{Gratier10}, which is roughly NE-SW.   
The infall velocity is of order $1 \kms$, which, although supersonic and thus capable of heating the gas, 
is not expected to create shocks strong enough to destroy grain mantles.  The large-scale velocity gradient due to 
the rotation curve is negligible.

Large-scale infall has thus far only been detected in DR21 \citep{Schneider10} 
although infall towards a number of cloud cores has been detected \citep[e.g.][]{chavarria10}, typically via inverse 
P-Cygni profiles.  
If we really are tracing infall at the scale of $10-20$ pc, we can estimate the accretion rate as a check.  
By measuring the emission to the SW and NE in respectively the lower right and upper left panels of 
Fig.~\ref{m33c_fig_mom0c}, we find that fluxes of 0.2 Jy$\kms$ and 0.15 Jy$\kms$ come from the apparently 
infalling gas, representing about 10\% of the total CO(1--0) flux from the cloud.
Assuming that the CO emission allows us to trace H$_2$ mass, some 4300~\msol\ may be falling onto the central GMC.  
For a distance of 20 pc and a velocity of $1 \kms$, the resulting rate is 
4300~\msol $/2 \times 10^7 {\mathrm yr} \approx 0.0002$ \msol/yr.  Such an accretion 
rate is highly plausible and an order of magnitude less than what was estimated for DR21.
The presence of two independent clouds cannot be totally excluded but they must either be much further apart than their projected distance or must have highly differing velocities perpendicular to the line of sight (i.e. not observable).

Averaged over the estimated cloud area of about $4000$ pc$^2$, the cloud surface density is really very low, some 11 \msol pc$^{-2}$.  However, this is an order of magnitude greater than the average stellar surface 
density so far from the center of M~33, so the presence of the GMC represents a real local center of gravity.
Presumably, this is what enables such weak inflow to be detected.


\bibliographystyle{aa}
\bibliography{jb}

\end{document}